# CASP: An evaluation dataset for formal verification of C code

Niclas Hertzberg[1][0009−0002−6048−7132], Merlijn Sevenhuijsen[2,3][0009−0002−1114−4395], Liv Kåreborn[1][0009−0004−8576−337X], and Anna Lokrantz[1][0009−0005−5826−1398]

[1] AI Sweden, Stockholm, Sweden
`{niclas.hertzberg, liv.kareborn, anna.lokrantz}@ai.se`
[2] Scania, Södertälje, Sweden
`{merlijn.sevenhuijsen}@scania.com`
[3] KTH Royal Institute of Technology, Stockholm, Sweden

**Abstract.** Recent developments in Large Language Models (LLMs) have shown promise in automating code generation, yet the generated programs lack rigorous correctness guarantees. Formal verification can address this shortcoming, but requires expertise and is time-consuming to apply. Currently, there is no dataset of verified C code paired with formal specifications that enables systematic benchmarking in this space. To fill this gap, we present a curated evaluation dataset of C code paired with formal specifications written in ANSI/ISO C Specification Language (ACSL). We develop a multi-stage filtering process to carefully extract 506 pairs of C code and formal specifications from The Stack 1 and The Stack 2. We first identify C files annotated with formal languages. Then, we ensure that the annotated C files formally verify, and employ LLMs to improve non-verifying files. Furthermore, we post-process the remaining files into pairs of C code and ACSL specifications, where each specification-implementation pair is formally verified using Frama-C. To ensure the quality of the pairs, a manual inspection is conducted to confirm the correctness of every pair. The resulting dataset of C-ACSL specification pairs (CASP) provides a foundation for benchmarking and further research on integrating automated code generation with verified correctness.

## 1 Introduction

Large Language Models (LLMs) for code generation have achieved remarkable results in recent years, showing strong performance on tasks such as generating syntactically valid functions and providing code completions [7,8,20,23]. While LLMs demonstrate value across many coding tasks, their utility remains limited in domains with strict safety and quality requirements, such as safety-critical systems, where software failures can lead to severe consequences. A key limitation is that code generated by LLMs cannot be reliably guaranteed to be correct.

In contrast to the probabilistic nature of LLMs, formal specification languages offer a potential solution to the nondeterministic behavior of LLMs. These formal specification languages provide robust means to specify program behavior, which can then be verified using formal verification tools. However, the adoption of such formal specification languages faces practical challenges: the pairs of formal specifications and associated code must be manually written, which is both time-consuming and requires expertise.

Given that manually creating such pairs is time-consuming, LLMs provide a promising application of being able to generate code from specifications or vice versa.

The task of generating specifications from code and code from specifications is arguably a most fundamental use case for applying LLMs to formal specification generation, directly targeting the most time-consuming tasks. Furthermore, like the related task of correcting non-verifying code, these tasks provide an easily interpretable success metric through formal verification since the generated pairs either verify or fail to verify.

However, evaluating the feasibility and progress of such an approach requires ways to measure the performance of LLMs in the form of dedicated evaluation datasets. We present such a dataset, consisting of verified pairs, that is specifically designed to benchmark this core generative capability.

Existing datasets containing ACSL specifications [1, 2, 4, 11, 25] and C code have two significant limitations. First, the datasets are limited in size and therefore lack the breadth needed to cover diverse real-world use cases. Lacking large and diverse datasets, researchers cannot draw general conclusions regarding LLMs' abilities to generate formally verified pairs.

Second, existing datasets are typically distributed as collections of whole C source files, where specifications are embedded as comments. This format requires a non-trivial parsing and extraction step to isolate individual specification-implementation pairs before they can be used in an evaluation pipeline for pair-generation tasks. These two limitations hinder progress in integrating formal methods into the software development process.

Our work addresses this fundamental gap by providing a dataset with sufficient breadth and volume to give researchers a reliable benchmark. Our dataset consists of C code and formal specifications in ANSI/ISO C Specification Language (ACSL), chosen for its adoption by the Frama-C [10, 18] verification platform, which is used in both academic research and industrial contexts for verifying critical properties of C programs [12, 27]. We call this dataset CASP, short for C-ACSL specification pairs. We create CASP by first sourcing all C files from *The Stack v1* and *The Stack v2* repositories. The C files are then filtered using three steps: (1) identifying and retaining high-quality C files containing formal specification in ACSL; (2) ensuring that the ACSL-annotated C files formally verify, and attempting to correct these if they do not verify (3) extracting individual function implementations and their formal specifications to obtain function-specification pairs.

This paper offers the following contributions:

1. We present CASP: a unique dataset of C code paired with associated formal specifications in ACSL, accessible at: `https://huggingface.co/datasets/nicher92/CASP_dataset`. The dataset contains 506 pairs. These pairs are systematically extracted from large-scale open-source datasets (The Stack v1 and v2) and are formatted as pairs in order to be amenable to LLM evaluation.
2. We share the complete files from which each pair was taken, accessible at: `https://huggingface.co/datasets/nicher92/CASP_source_files`. The number of files is much larger than previously available public datasets, offering significantly more data for training and evaluation. Additionally, the files – and by extension the pairs– are all "minimally complete" – meaning the files have no dependencies other than the standard C libraries.
3. We detail our filtering, verification, and post-fixing procedures, ensuring that each file, as well as each code-specification pair, formally verifies and is consistent with one another.

This dataset fills a gap by providing a valuable resource for benchmarking and training LLMs on the task of specification generation from code and vice versa. By offering a conveniently formatted dataset consisting of verified, minimally complete pairs of specifications and associated code, our work supports the development of advanced tools for software verification by contributing to the creation of more reliable software systems.

The rest of this paper is organized as follows. We first provide a brief overview of formal verification, in particular the ANSI/ISO Specification Language (ACSL) in Section 2. Then in Section 3, we review existing formal specification datasets and their limitations. Section 4 explains our data collection methodology, followed by our file verification process in Section 5. We then describe how we divided the files into specification-function pairs Section 6. Section 7 presents the composition and key statistics of the dataset, with a discussion and analysis in Section 8. Finally, Section 9 presents our conclusions and suggests directions for future work.

## 2 Background

This section provides background on the specification language and verification tools used in our dataset. In particular, we describe the ANSI/ISO C Specification Language (ACSL) and the Frama-C verification framework, with a focus on the WP and RTE plugins used to check correctness and runtime safety.

In the dataset, we focus on ANSI/ISO C Specification Language (ACSL), which enables the formal verification of C code. The language is designed for use with the Frama-C verification framework, a framework for static analysis and deductive verification [3].

### 2.1 ANSI/ISO C Specification Language

ACSL is a contract-based specification language that allows formal verification of C programs by defining preconditions, postconditions, invariants, and memory access constraints. It is designed to be used with the weakest precondition plugin of Frama-C.

```c
/*@
    requires \valid(x) && \valid(y);
    assigns *x, *y;
    ensures *x == \old(*y) && *y == \old(*x);
*/
void swap(int* x, int* y) {
    int temp = *x;
    *x = *y;
    *y = temp;
}
```

**Fig. 1.** ACSL specification and associated implementation in C for a function swapping two integers.

Figure 1 demonstrates an ACSL-annotated function that swaps the values of two integer pointers. The `requires` clause (line 2) specifies that both variables x and y must be valid pointers before execution. The `\valid` predicate ensures that the pointers reference accessible memory. The `assigns` clause (line 3) explicitly states that the function modifies the memory locations pointed to by x and y, making side effects explicit. The `ensures` clause (line 4) guarantees that after execution, the values of variables x and y have swapped. The `\old` keyword refers to the values before function execution, ensuring that the function correctly swaps the values.

### 2.2   The Frama-C framework

Frama-C is a modular analysis framework for C programs that supports a variety of verification techniques, including runtime error detection and deductive verification. In this work, we use two of its key plugins: WP and RTE.

The `WP (Weakest Precondition)` plugin generates proof obligations called goals from ACSL-annotated C code using weakest precondition calculus. These obligations are passed to SMT solvers (e.g., Alt-Ergo, Z3, CVC4), which attempt to automatically prove that these goals within a given timeout and number of steps.

The `RTE (Runtime Error)` plugin instruments the program with ACSL annotations that check for common runtime errors, including division by zero, null pointer dereference, invalid memory access, and integer overflow. The WP plugin then verifies these additional checks as part of the deductive verification process. Together, WP and RTE enable Frama-C to verify both functional correctness and runtime safety.

## 3   Related Work

This section reviews prior work relevant to our dataset, including large-scale source code collections, existing datasets containing formal specifications in C, and recent efforts to combine ACSL with automated code generation using LLMs.

### 3.1   Large-Scale Source Code Datasets

The availability of large source code datasets is fundamental for training and evaluating large language models (LLMs) on code-related tasks. Notably, The Stack v1 [19] and v2 [22], created as part of the BigCode Project[4], provide vast repositories of permissively-licensed source code across numerous programming languages. The Stack v1 comprises approximately 546 million files totaling 6.4 TB, covering 358 programming languages. The subsequent release, The Stack v2 , significantly expanding this collection to over 3 billion files (67.5 TB) in more than 600 languages, further enhancing the diversity and volume available for model training and evaluation.

|  | The Stack v1 | The Stack v2 |
|---|---|---|
| Total number of files | 5.46M | 3B |
| Number of C files | 19.88M | 40.88M |

**Table 1.** Total number of files and number of C files in The Stack v1 and v2. Note that M refers to million, and B refers to billion.

### 3.2   Formal Specification Datasets used in Literature

Existing collections of C code annotated with ACSL specifications primarily originate from research projects, serve as educational materials, or have uncertain origins.

---
[4] https://www.bigcode-project.org/

| Dataset source | ACSL annotated C files | Minimally complete verified programs |
|---|---|---|
| Frama-C-problems [25] | 51 | 9 |
| X509-parser [1] | 6 | 0 |
| Verker [14] | 48 | 1 |
| ACSL By Example [15] | 86 | 3 |
| WP examples [5] | 295 | 134 |
| ACSL proved [13] | 34 | 10 |
| VecoSet [4] | 15 | 14 |

**Table 2.** Existing Formal Specification datasets for C code.

Datasets developed in research contexts often function as case studies for formal verification techniques [15, 25] or as benchmarks for evaluating analysis tools [4], typically created through manual ACSL annotation of C code. A significant educational resource is the ACSL tutorial [5], designed to teach specification writing through hands-on exercises. This tutorial contains numerous examples, many intentionally left incomplete for learners to finish, reflecting its pedagogical goal.

Common characteristics of these available datasets include small size, formatting as individual files (sometimes with dependencies to other files like .h headers), and a structure tailored to their specific origin or teaching objective rather than forming a larger corpus designed for evaluating LLMs.

### 3.3   Previous Work on LLMs for C Code and ACSL Specifications

Prior research has explored the intersection of C programming and ACSL specifications, particularly in the context of leveraging LLMs for code generation and verification. Minal et al. [24] investigated the feasibility of using LLMs to generate automotive safety-critical embedded C code from both natural language and ACSL specifications. Their study demonstrated the potential of producing compilable and partially verifiable code without iterative backprompting or fine-tuning, though the limited scope of their case studies highlighted the need for more extensive datasets. Similarly, Sevenhuijsen et al. [26] developed a tool that employs a two-step process of initial code generation followed by iterative improvement using feedback from compilers and formal verifiers. The tool successfully generated verified C programs for a majority of the problems in their benchmark set, underscoring the effectiveness of combining formal specifications with automated code generation. However, the relatively small number of code samples in these studies indicates a pressing need for larger, more comprehensive datasets to draw stronger conclusions and enhance model performance.

Similar work has begun to infer ACSL specifications automatically from C code. Granberry et al. prompt GPT-4 with source code plus test inputs and static-analysis warnings, then refine the output of the model until it verifies in Frama-C [16]. Wen et al. apply heuristic post-processing to GPT-4 predictions, correcting syntax and adding safety clauses so the resulting contracts verify more reliably [28]. Together, these studies show that coupling large language models with formal-methods feedback is a promising route to automatic ACSL specification generation.

## 4   Dataset Collection

Our methodology for collecting and curating a dataset consisting of C functions paired with their corresponding ACSL specifications. We employ a three-step data collection process, shown in Fig-ure 2. This section focuses on the first step of our dataset creation, where we gather a large collection of source files and then iteratively filter the files in order to isolate ACSL-annotated C files without other dependencies.

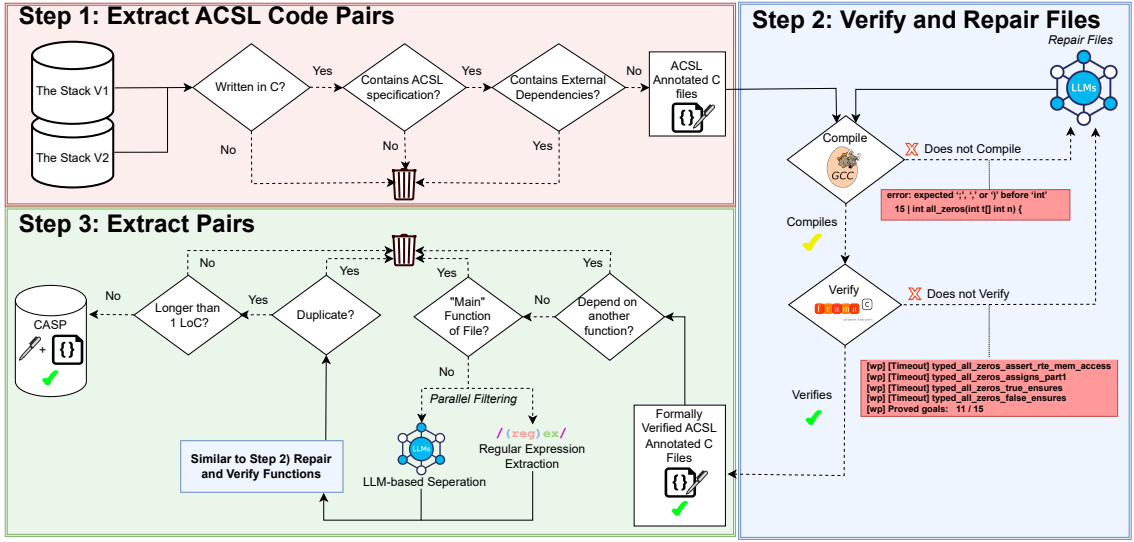

**Fig. 2.** Overview of the three-step dataset construction pipeline. `Step 1` involves using files from The Stack v1 and v2, which are filtered to identify C files containing ACSL specifications without external dependencies. Step 2 compiles and verifies each annotated file, and files that fail this are automatically repaired using LLMs and re-evaluated. Step 3 transforms successfully verified files into minimal specification-implementation pairs and includes them in CASP.

### 4.1   Downloading The Stack

As an initial step in our data processing pipeline, all files tagged as being written in `C` were downloaded from the deduplicated versions of Stack V1 and V2. Table 1 shows the total number of files in The Stack (V1 and V2) and the subset identified as C files.

### 4.2   Extracting Files containing ACSL specifications

After our initial collection of files, we applied regular expression filters in several steps, in order to extract files that contain ACSL-like annotations. The regular expression patterns were authored by a formal methods expert and are detailed in Appendix A. Each pattern is associated with a *confidence label* indicating whether it can appear in multiple formal languages, such as VeriFast [17] (*possible overlap*), or is unique to a specific language (*exclusive*). We iteratively applied stricter quality filters, only keeping files containing specific ACSL syntax, which left us with 2958 files, as can be seen in Table 3.

| Processing Step | Count | Description |
| --- | --- | --- |
| *Initial Data Collection* | | |
| Stack 1 (deduplicated) | 8,625,559 | Raw code samples |
| Stack 2 (deduplicated) | 17,093,668 | Raw code samples |
| Combined total | 25,719,227 | Total initial code samples |
| *Dataset A Creation* | | |
| Initial regex filtering | 14,525 | Files matching basic patterns |
| Stricter filtering | 5,916 | Files with formal specifications |
| ACSL filtering | 2958 | Files with ACSL |
| Standard/No import filtering | 1180 | Minimally complete files |

**Table 3.** Dataset collection and filtering process. Starting from The Stack (deduplicated versions 1 and 2), we progressively filter files to identify those containing ACSL specifications. The strict filtering uses the pattern `/\*@.*?(predicate|requires|ensures).*?\*/`.

### 4.3 Minimally complete C files

Many of the collected C files depended on code from other files or non-standard libraries. These dependencies were often complex, making the extraction of verifiable functions and specifications challenging.

To address this issue, we applied an additional filtering step: we only retained files that are self-contained, with dependencies limited to standard C libraries. We refer to these as "minimally complete files" since the specifications and functions they contain can be analyzed independently without requiring external code.

After this stage of our pipeline, we retained 1180 minimally complete C files (see Table 3) that contained ACSL specifications.

## 5 Verifying and Curating CASP source files

This section describes the second step of our three-step process depicted in Figure 2. Specifically, it describes our method for verifying the correctness of our minimal complete files and our attempts to correct files that do not compile or formally verify.

### 5.1 Method

To formally verify the minimally complete C files, we attempt to verify if the C implementation meets the formal specifications in these files. The verification process is done by two plugins of Frama-C [9], which we described in Section 2.2. We verified each source file using Frama-C version 30.0 (Zinc) with the WP and RTE plugins. The verification was performed using multiple SMT solvers to combine their strengths: Z3 version 4.8.12, Alt-Ergo version 2.6.0, and CVC4 version 1.8. We configured the WP plugin with a 500,000-step limit and a 60-second timeout per proof goal.

For each source file, we either successfully completed all of the goals set by the WP and WP-RTE [6] plugins of Frama-C, or we retrieved the non-verifying goals from Frama-C and attempted to repair the specification. Files that fail to meet the specifications were sent to an LLM (Gemini 2.0 Flash) along with the non-verifying goals from Frama-C and a prompt requesting to update the code such that all goals are verified. We then iteratively attempted to correct each failing file for a maximum of seven iterations. Through this process, we ended up with 469 verified files in our final dataset.

| Category | Count |
|---|---|
| Minimally complete files | 1180 |
| Minimally complete files verified without modifications | 292 |
| Minimally complete files verified with modifications | 177 |
| **Total verified minimally complete Files** | 469 |

Table 4. File Analysis Summary

**Prompt Engineering** Our approach to prompt design was iterative, refining the instructions for the LLM based on patterns of verification errors observed in its outputs. We focused particularly on addressing common verification challenges, such as proper contract clause ordering, memory access specifications, and strategic assertion placement to guide proofs. The LLM was prompted to make minimal changes to the code, and also to output what it "thinks" the user intended with their code in order to limit deviation from the original code. The complete prompt used in our processing pipeline can be found in Appendix B.

## 6  CASP pair creation

This section explains step three of the three-step process mentioned in Figure 2. It describes the means for separating the verifying files into specification implementation pairs.

### 6.1  Motivation for Specification-Implementation Pairs

Creating verified formal specifications and function pairs offers three advantages over verified C files. First, using specification-function pairs provides a more decoupled method for evaluating LLM performance on formal verification tasks than using entire files. For example, since each specification corresponds to a function implementation, it is possible to assess the generative capabilities of an LLM given a formal specification. Second, these pairs ensure that the specification is logically consistent with the implementation and practically implementable. This addresses a fundamental challenge in formal methods where abstract specifications may contain logical inconsistencies or unrealizable requirements. Third, the structure of the dataset supports bidirectional evaluation — from specification to code and vice versa – which in turn supports a broader range of research questions related to the generative capabilities of LLMs.

### 6.2  Minimally complete files to minimally complete pairs

To create minimally complete pairs from minimally complete files, we selected function implementations according to the following requirements:

- The functions do not depend on other functions in a file.
- The functions are not main functions.

The decision to focus on standalone functions was guided by two primary factors. Methodologically, it creates a constrained test that directly evaluates an LLM's core ability to translate between a specification and an implementation. Practically, the task of identifying, extracting, and verifying the complete context for functions with numerous dependencies from large codebases is often intractable. Our approach therefore ensures that each pair in CASP is a self-contained and verifiable unit.

We utilized two parallel pipelines in order to extract pairs of ACSL specification and C implementation from our source files: One based on utilizing an LLM and one based on regular expressions. Each minimally complete file that fulfilled the above requirements was sent to both pipelines.

The regular expression-based pipeline comprises three steps: first, extracting function implementations, second, their corresponding ACSL specification if present, and third, any additional dependencies needed for verification of the pair. Any functions without associated specifications were removed.

The LLM-based pipeline (Using Gemini 2.0 Flash) consisted of prompting the model to extract functions, their associated specification, and any additional dependencies [5]. Regular expression for function extraction and prompt can be found in Appendix C.

We take the union from the resulting pairs from both pipelines, which were then verified by Frama-C, similarly as described in 5.1; any remaining unverified pairs were manually post-fixed.

The resulting union of pairs, 513 in total, from the two pipelines was then filtered in two steps: We performed exact deduplication of the C implementations and only kept function implementations longer than one line of code, leaving us with 506 pairs.

## 7 Dataset statistics

In this section, we provide a statistical overview of our dataset. First, in Fig. 3 we show the length distribution measured as lines of C code for our pairs, and find that most of the CASP pairs are short to medium in length. Correspondingly, the number of lines of ACSL for each pair can be seen in the distribution plot in Fig. 4, most specifications are short to medium in length, with some more complex outliers.

Beyond characterizing the CASP pairs themselves, we also analyze the diversity and novelty of the verified C source files from which these pairs were derived. Comparing the source files to existing file-based ACSL datasets allows us to assess the breadth of our data collection. Since our dataset is collected from open-sourced code, there is a possibility of overlap between CASP and other open-sourced datasets containing ACSL. In order to measure potential overlap, we embed our files and compare the semantic similarity of our files to existing datasets. The comparison is done in two ways: first, using a t-SNE plot in two dimensions, and second, by a nearest neighbour comparison. An in-depth analysis of the semantic distributions can be found in Section 8.2.

### 7.1 Semantic distribution of file contents

We downloaded existing datasets (see Table 2) containing ACSL specifications and C code. We filtered each C file in all datasets – including the CASP source files – so that all files were minimally complete, ensuring a fair comparison between datasets. We then embedded the files that verify using CodexEmbed [21] – a model specifically developed for code retrieval. We used the 2B parameter model variant. The embedding model has a maximum context length of 4096, which means that longer code samples were truncated, potentially affecting their representation.

---

[5] With additional dependencies we mean standard imports, type and variable declarations, logic predicates, etc.

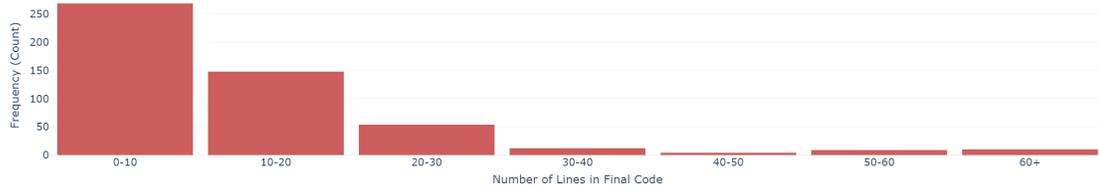

**Fig. 3.** Distribution of lines of C code for each CASP pair, excluding specification and imports. Most programs are short to medium in length. The X axis indicates lines of code, and the Y axis indicates a number of occurrences. Outliers over 60 total lines of code are binned together.

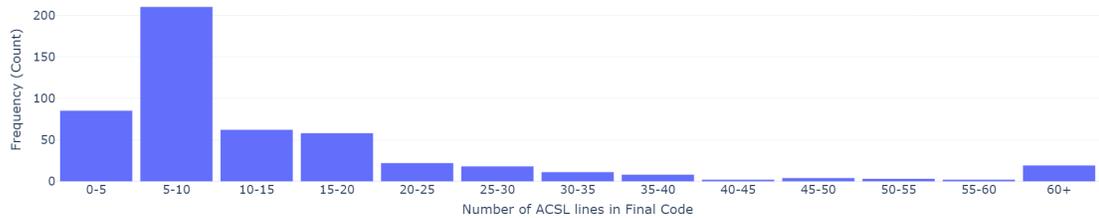

**Fig. 4.** Distribution of total lines of ACSL for each CASP pair. The X axis indicates the total number of lines of ACSL, and the Y axis indicates the number of occurrences. Outliers over 60 total lines of ACSL are binned together.

Using these high-dimensional embeddings, we visualized the semantic relationships between files using t-SNE (Fig. 5) and quantitatively analyzed the similarity distribution by calculating nearest neighbor distances (Fig. 6).

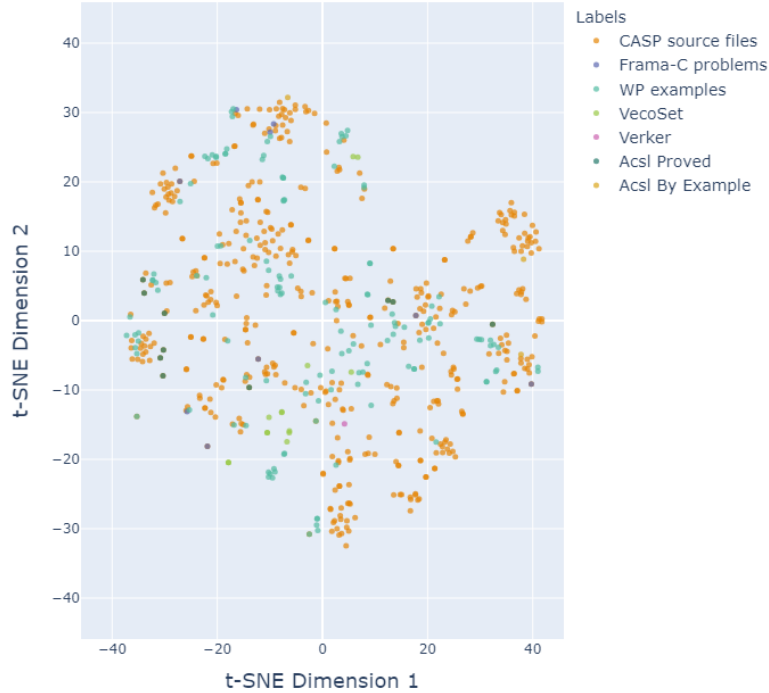

**Fig. 5.** t-SNE visualization of embeddings from various datasets that verifies without any external imports (CASP source files, Frama-C problems, VecoSet, etc.). The plot shows the projection of high-dimensional embeddings into a 2D space, where proximity suggests similarity. Colors indicate the source dataset as shown in the legend. Any difference in embeddings indicates a difference in file content.

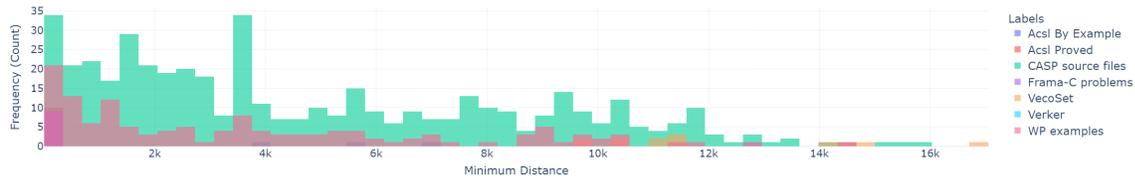

**Fig. 6.** Nearest neighbours of various datasets (CASP source files, Frama-C problems, VecoSet, etc.). The plot shows the distance between each embedded source file, where the source file is from and how similar its nearest neighbour is. A dataset containing a wide variety of files would contain more points to the right on the x axis and vice versa.

## 8   Discussion and Analysis

In this section, we describe our findings on the current state of openly sourced, ACSL-annotated code. We then provide an analysis of CASP and the source files from which CASP was derived. Finally, we discuss our method for LLM-based specification repair along with implications for verification and specification generation.

### 8.1   Current State of Affairs and the Need for CASP

Our investigation confirmed a significant challenge for researchers in automated verification and specification generation: the pronounced scarcity of openly accessible C code accompanied by ACSL annotations. Furthermore, where such code exists, a substantial portion exhibits quality issues, often failing verification by Frama-C. This scarcity presents challenges for researchers seeking to build comprehensive datasets for LLM training or benchmarking purposes. We hypothesize that the main reason for this scarcity is that while ACSL is a fairly standard verification language, much of the code where it is present is not openly available on GitHub with permissive licensing, and is therefore not included in The Stack 1 and 2.

Despite this scarcity, CASP is the largest openly released dataset containing ACSL specifications and C code so far. Additionally, the dataset is formally verified and formatted in a way conducive to evaluating LLMs. Another key strength of CASP lies in its inherent diversity. Since the collected code samples were authored by numerous different programmers, they exhibit considerable variety in implementation styles, algorithmic approaches, and specification patterns. This diversity strengthens the utility of our dataset for various research applications, as it represents a broad spectrum of real-world specification practices rather than the more uniform patterns that might emerge from a single team or project.

### 8.2   Dataset Composition and Analysis

The embedded source files from CASP – when visualized using t-SNE in Fig. 5 – occupy a broader range of the semantic space and therefore show greater diversity than existing datasets, encompassing most regions where samples from existing datasets are located. We hypothesize that there are two reasons for this: The CASP source files are substantially more numerous than existing datasets, and the source files originate from multiple sources and numerous different authors. It should be noted that the ACSL annotated C files from other datasets often contain imports from .h files – which we do not include – causing many of the files to not verify.

Furthermore, our analysis (see Fig. 6) reveals that several files from the datasets Frama-C Problems, ACSL Proved, WP Examples, and CASP are similar to at least one other file in one of the datasets. Beyond these clusters of similarity, we found a broad distribution of datapoint relationships across the similarity spectrum. Overall, we find that there is fairly limited overlap between CASP and previous datasets, since only approximately 35 CASP source files are very similar to some other file in any of the other datasets. One explanation for this limited overlap is that we remove files that are dependent on files not found in the standard libraries, including .h files. Many of the C files in the existing datasets contain .h files and are, therefore, naturally filtered if found.

### 8.3 LLM-Based Specification Repair and Pair Extraction

While not the primary focus of this work, our approach to repairing faulty ACSL-annotated C files using LLMs showed meaningful success.

Our methodology successfully corrected 177 files out of the 888 that initially failed verification, representing a 19.9% success rate among files requiring modification. The success rate highlights the challenging nature of formal specification repair, even for advanced LLMs. Nevertheless, the fact that nearly one-fifth of problematic specifications could be automatically corrected suggests potential for improvement in this area. One explanation for this rate of success is related to the limited amount of ACSL-annotated C code that is openly available: it is difficult to train an LLM to understand the syntax and semantics of a formal language that the LLM has barely encountered previously. Another reason could be that the non-repairable files themselves were poorly written and therefore difficult for an LLM to repair, especially given the constraint not to deviate from the original implementation.

In addition to repairing ACSL-annotated C files, we also utilized LLMs for function extraction. We found that LLMs struggle in some cases but can be complemented by straightforward rule-based methods based on regular expressions. This combined approach proved to be more successful than using either approach on its own.

## 9  Conclusions and Future Work

This paper has demonstrated our approach to creating CASP: a dataset consisting of 506 verified and deduplicated C code functions paired with ACSL specifications.

CASP has only been publicly available since June 2025, it serves as a timely and uncontaminated benchmark for all current models and any future models with a training data cutoff before this date, but future work should also address the long-term maintenance of CASP as a benchmark. To mitigate the risk of data contamination from future model training, a portion of the dataset could be reserved for a private test set, and our methodology could be used to generate new versions of the benchmark over time.

Additionally, our findings suggest several promising directions for extending this work. We suggest two main avenues of exploration, extending CASP and specific dataset applications.

A natural extension could involve exploring different data sources to expand our data set. For example, academic papers and technical documentation often contain high-quality specifications created by experts that could yield additional examples of formally verified pairs. Synthetic data generation, using CASP for seed prompts with modifications to promote greater diversity in the generated code samples, is another promising avenue to explore.

As CASP was created first and foremost with evaluation of LLMs in mind, a natural next step in terms of applications would be to evaluate a wide variety of different LLMs on generating code from formal specifications and vice versa. Future work, possibly following an extension of the dataset, might also explore training specialized models for formal verification tasks or developing automated tools for specification generation and repair.

## A   Regex patterns used

Table 5. Patterns for ACSL, and Verifast Annotations

**ACSL**

```
/@(?:(?!@/)[\s\S])*?\brequires\b(?:(?!@/)[\s\S])*?/
/@(?:(?!@/)[\s\S])*?\bensures\b(?:(?!@/)[\s\S])*?/
/@(?:(?!@/)[\s\S])*?\bassigns\b(?:(?!@/)[\s\S])*?/
/@(?:(?!@/)[\s\S])*?\binvariant\b(?:(?!@/)[\s\S])*?/
/@(?:(?!@/)[\s\S])*?\baxiomatic\b(?:(?!@/)[\s\S])*?/
/@(?:(?!@/)[\s\S])*?\blemma\b(?:(?!@/)[\s\S])*?/
/@(?:(?!@/)[\s\S])*?\bpredicate\b(?:(?!@/)[\s\S])*?/
/@(?:(?!@/)[\s\S])*?\blogic\b(?:(?!@/)[\s\S])*?/
/@(?:(?!@/)[\s\S])*?\bbehavior\b(?:(?!@/)[\s\S])*?/
/@(?:(?!@/)[\s\S])*?\bdisjoint behaviors\b(?:(?!@/)[\s\S])*?/
/@(?:(?!@/)[\s\S])*?\bcomplete behaviors\b(?:(?!@/)[\s\S])*?/
/@(?:(?!@/)[\s\S])*?\bassumes\b(?:(?!@/)[\s\S])*?/
//@\s*\brequires\b
//@\s*\bensures\b
//@\s*\bassigns\b
//@\s*\binvariant\b
//@\s*\baxiom\b
//@\s*\blemma\b
//@\s*\bassert\b
loop invariant
loop assigns
loop variant
\\old\b
\\at\b
\\nothing\b
\\max\b
\\min\b
\\result\b
\\forall\b
\\exists\b
\\sum\b
\\sizeof\b
\\valid\b
\\valid_read\b
\\is_finite\b
```

**Verifast**

```
/@(?:(?!@/)[\s\S])*?\bopen\b(?:(?!@/)[\s\S])*?@/
/@(?:(?!@/)[\s\S])*?\brequires\b(?:(?!@/)[\s\S])*?@/
/@(?:(?!@/)[\s\S])*?\bensures\b(?:(?!@/)[\s\S])*?@/
/@(?:(?!@/)[\s\S])*?\bassert\b(?:(?!@/)[\s\S])*?@/
/@(?:(?!@/)[\s\S])*?\bfold\b(?:(?!@/)[\s\S])*?@/
/@(?:(?!@/)[\s\S])*?\bunfold\b(?:(?!@/)[\s\S])*?@/
/@(?:(?!@/)[\s\S])*?\blemma\b(?:(?!@/)[\s\S])*?@/
/@(?:(?!@/)[\s\S])*?\bpredicate\b(?:(?!@/)[\s\S])*?@/
/@(?:(?!@/)[\s\S])*?\bopen\b(?:(?!@/)[\s\S])*?@/
/@(?:(?!@/)[\s\S])*?\bclose\b(?:(?!@/)[\s\S])*?@/
/@(?:(?!@/)[\s\S])*?\binvariant\b(?:(?!@/)[\s\S])*?@/
/@(?:(?!@/)[\s\S])*?\bpointer\(\b(?:(?!@/)[\s\S])*?@/
/@(?:(?!@/)[\s\S])*?\bmalloc_block(?:(?!@/)[\s\S])*?@/
//@\s*\binclude\b
//@\s*\brequires\b
//@\s*\bensures\b
//@\s*\bassert\b
//@\s*\bfold\b
//@\s*\bunfold\b
//@\s*\binvariant\b
//@\s*\blemma\b
//@\s*\bopen\b
//@\s*\bclose\b
//@\s*\bleak\b
```

## B  Gemini Prompt

The following is the prompt used to instruct Gemini to correct ACSL specifications in C code based on Frama-C error messages.

**Listing 1.1.** Main prompt given to Gemini

```
Your task is to correct ACSL specifications based on C code and error messages from Frama-C.
Your goal is to repair the ACSL specifications so that they pass Frama-C's verification.
Do not alter the C code unless absolutely necessary.
Focus on correcting ACSL specifications to address common errors such as:

Invalid ACSL syntax
Type mismatches in ACSL expressions
Loop invariants that are not strong enough or incorrect
Precondition or postcondition failures
Memory access errors or incomplete memory specifications
Incomplete or incorrect assigns clauses
Timeout issues in proof obligations

Pay special attention to:

Using precise memory specifications: \valid, \valid_read, \separated as appropriate
Ensuring loop invariants are strong enough to prove postconditions
Adding explicit loop assigns clauses to clarify what loops modify
Using complete behaviors and disjoint behaviors when appropriate
Adding strategic assertions to guide the prover
Using \exists and \forall quantifiers correctly
Ensuring that array bounds are properly specified
Not to add undefined variables that are not defined in the code previously.
Loop assigns is not allowed after loop variant so they need to be prior to the loop variant
Wrong order of clause in contract: behavior should be before complete or disjoint for example
Using correct syntax for behaviors: each behavior should be declared separately using the
      behavior keyword, not enclosed in braces; complete behaviors and disjoint behaviors should
      be followed by a comma-separated list of behavior names without braces
The ACSL specifications for a function should be above the function, not below it.
Avoid adding main() functions if not present in the original code.
In general the changes should not attempt to alter the purpose of the original code.

For timeout issues, consider:

Simplifying complex specifications
Breaking down properties into smaller, more provable assertions
Using different specification styles (direct ensures vs. behaviors)
Adding intermediate assertions to guide the proof

Output the corrected file in JSON format, including a brief explanation of the changes made and
      any assumptions.

Input:
C Code (Previous Attempt):
{file_content}
Frama-C Error Message (From Previous Attempt):
{error_message}
```

**Listing 1.2.** Context prompt for subsequent iterations

```
Context from Previous Gemini Attempt (that produced the code above):
Previous Explanation: {prev_explanation}
Previous Assumptions: {prev_assumptions}
Previous Strategies Suggested: {prev_strategies}
Based on the previous attempt's code, the resulting error message, and the previous explanation/
      assumptions, please refine your corrections to address the remaining issues. Focus
      specifically on the errors highlighted in the Frama-C message.
```

Listing 1.3. Context prompt for first iteration

```
This is the first attempt to fix the provided code and error message in this refinement cycle.
    Please analyze the code and error carefully.
```

Listing 1.4. Required output format

```
Output Format:
Please provide your response in the following JSON format:
{
  "explanation": "Explanation of changes made in this attempt",
  "assumptions": "Any assumptions made during this correction process",
  "fixed_code": "Complete corrected code with fixed ACSL specifications here",
  "strategies": "Suggestions for prover strategies if timeout issues persist (e.g., specific
      provers, timeouts, steps)"
}
```

## C  Function pair extraction

Listing 1.5. Regex for extracting non main functions

```
        r'''
        (?P<signature>
            (?:[a-zA-Z_][\w\s\*\(\),]*?)    # return type and qualifiers
            \s+                             # whitespace
            (?!(main)\s*\()                 # not 'main'
            [a-zA-Z_]\w*                    # function name
            \s*\([^;]*?\)                   # parameter list
        )
        \s*\{                               # opening brace of body
        '''
```

Listing 1.6. Gemini prompt for function and specification pair extraction

```
You are given a C source file that contains one or several functions with corresponding ACSL
    specifications and additional dependencies.

Your task is to extract all functions that are independent of other functions, except 'main',
    which should be excluded. A function is considered independent if it does not call or rely
    on other user-defined functions in the file.

For each such function:
1. Extract the full function implementation (signature and body).
2. Extract the ACSL specification that precedes it (typically marked by /*@ or //@).
3. Identify and include only the minimal dependencies required for Frama-C verification of the
    function. This may include:
    - '#include' directives (e.g., '<stdbool.h>')
    - '#define' macros
    - global constants or variables used in the function

Return a JSON object for each function with the following fields:
- "function_implementation": the code of the function (not 'main')
- "acsl": the ACSL specification
- "dependencies": the minimal required includes/defines/globals

If a function depends on another user-defined function in the same file, skip it.

C source code:
{file_content}
```